\documentclass[twocolumn,aps,prd,amsmath,amssymb,nofootinbib]{revtex4-1}\newcommand{\preprintnumber}{\hfill MIT-CTP 4551\,\,\,\,\maketitle}

\usepackage{graphicx,bm}

\newcommand{\beq}{\begin{equation}}
\newcommand{\bea}{\begin{eqnarray}}
\newcommand{\eeq}{\end{equation}}
\newcommand{\eea}{\end{eqnarray}}

\newcommand{\mpl}{M_{Pl}}
\newcommand{\ms}{\gamma}
\newcommand{\lm}{\Lambda}

\begin{document}

\title{Inflation Driven by Unification Energy}

\author{Mark P.~Hertzberg$^{1,2,*}$ and Frank Wilczek$^{1,\dagger}$}
\affiliation{$^1$Center for Theoretical Physics and Dept.~of Physics, Massachusetts Institute of Technology, Cambridge, MA 02139, USA\\
$^2$Institute of Cosmology, Department of Physics and Astronomy,  Tufts University, Medford, MA 02155, USA}

\date{\today}

\begin{abstract}
We examine the hypothesis that inflation is primarily driven by vacuum energy at a scale indicated by gauge coupling unification.  Concretely, we consider a class of hybrid inflation models wherein the vacuum energy associated with a grand unified theory condensate provides the dominant energy during inflation, while a second ``inflaton'' scalar slow-rolls. We show that it is possible to obtain significant tensor-to-scalar ratios while fitting the observed spectral index. 
\end{abstract}

\preprintnumber

\section{Introduction}

\let\thefootnote\relax\footnotetext{$^*$Electronic address: {\tt mphertz@mit.edu}\\$^\dagger$Electronic address: {\tt wilczek@mit.edu}}

Cosmological inflation \cite{Guth:1980zm,Linde:1981mu,Albrecht:1982wi} provides an explanation for the large scale homogeneity and isotropy of the universe.  It also suggests a natural origin for small inhomogeneities, tracing them to spontaneous fluctuations in quantum fields.  Its qualitative and semi-quantitative predictions agree well with recent observations, notably including the approximately, but not precisely, scale-invariant spectrum of density perturbations.  Gravitational waves, resulting from quantum fluctuations in the metric field, are another possible consequence of inflation. The amplitude with which these occur, however, is sensitive to the energy density during inflation.  Within the present state of understanding, that energy density appears as an essentially free parameter.

In this letter we explore the possibility that vacuum energy density associated with grand unification \cite{Georgi:1974sy,Georgi:1974yf} dominates the energy density during inflation. Then the amplitude of primordial gravity waves is set, at least semi-quantitatively, by microphysics -- specifically, the energy scale of unification of gauge couplings \cite{Dimopoulos:1981yj}. 

\begin{figure}[b]
\includegraphics[width=\columnwidth]{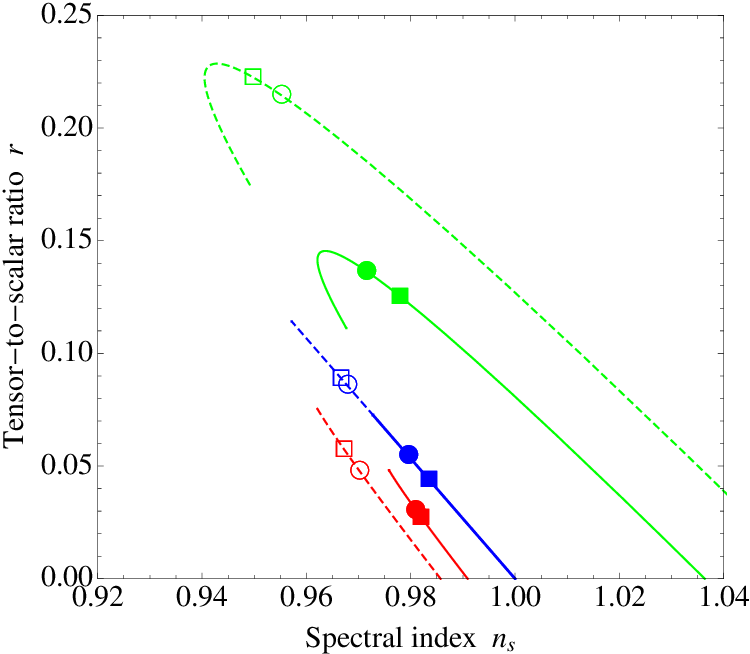}
\caption{Predictions for the spectral index $n_s$ and tensor-to-scalar ratio $r$ as we vary $u_d=U(\phi_*)/V_0$ for potentials that during inflation are monomials $U\propto|\phi|^p$. $u_d\to 0$ towards the lower right while $u_d\to\infty$ towards the upper left. Blue, red, and green curves are $p=1, 2/3, 3/2$ respectively.  Solid curves are for $N_e=55$ and dashed curves are for $N_e=35$ (see Section \ref{Modified}). The circles indicate the point $r_{\bigcirc}$ at which $u_d=1$ and the squares indicate the point $r_{\square}$ at which $\phi_*=1/\sqrt{G_N}$.}
\label{fig:nsvsr}
\end{figure}

We find that models implementing this idea can be consistent with existing data \cite{WMAP,Planck,BICEP}, and allow for a significant tensor-to-scalar, such as might be discovered in various upcoming searches. Our main results are summarized in Figure \ref{fig:nsvsr}, where we compute the predictions for the spectral index and tensor-to-scalar ratio in different parameter regimes.

In order to obtain the requisite number of e-foldings, we will need to introduce a scalar field $\phi$ that undergoes Planck-scale displacement during inflation.  Because of this, the global shape of its potential is sensitive to effects of quantum gravity and other corrections that are not well controlled theoretically.   We will discuss the resulting uncertainties as they arise.

\section{Model}\label{Unification}

To implement the idea that unification energy density drives inflation, we consider a Higgs field $h$ with potential
\beq
V_h = {\lambda\over 4}(h^2-v^2)^2
\eeq
where $v$ is the vacuum expectation value associated with gauge unification, and $\lambda$, taken to be $\mathcal{O}(1)$, is a self-coupling.  We have fixed the vacuum energy after symmetry breaking to be negligible.  
Unification of gauge couplings suggests $v\sim 10^{16}\,$GeV (in MSSM, see e.g., \cite{Carena:1993ag}), a value also broadly consistent with neutrino masses $m_\nu\sim 0.1$\,eV, generated via the see-saw mechanism.  The unification scale probably cannot be significantly lower, since unification brings in particles capable of mediating proton decay, and those must be very heavy.    
The vacuum energy stored in the field at the symmetric point $h=0$ is then
\beq
V_0 = {\lambda\over4}v^4\sim (10^{16}\,\mbox{GeV})^4
\eeq
We will investigate whether we can sit near that point in field space and drive inflation using $V_0$.

The minimal possibility, using the evolution of $h$ itself to drive inflation, is problematic.  Indeed, the (tachyonic) mass parameter $m_h\sim 10^{16}\,$GeV is significantly larger than the Hubble parameter $H = \sqrt {V_0 / 3} /  \mpl$, 
where $\mpl\equiv1/\sqrt{8\pi G}$ is the reduced Planck mass. Therefore the field will evolve away from $h = 0$ rapidly, and we will not have an extended period of inflation. 

This leads us to consider a form of hybrid inflation \cite{lindehybrid}. (Earlier analyses, often focused on low scale models, appear in \cite{sol1,sol2,lindeBH,carr1,carr4,lyth2,sfak}). We will consider the two field potential  
\beq
V(\phi,h) = {\lambda\over 4}(h^2-v^2)^2 + {g\over2}\phi^2 h^2+U(\phi)
\label{FullPot}\eeq
Now, because of the coupling term $\sim g\,\phi^2h^2$,  $h$ is stable around $h=0$ for large $\phi$. 

Our main freedom lies in the choice of potential $U(\phi)$.   During the first phase of inflation,  when $h=0$, the $\phi$ field evolves in the potential
\beq
V_\phi = V_0 + U(\phi)
\eeq

\section{Observables}\label{Observables}

Inflationary predictions are encoded in the slow-roll parameters
\bea
\epsilon(\phi) &=&  {\mpl^2\over 2}\left(V_\phi'\over V_\phi\right)^{\!2}\\
\eta(\phi) &=&  \mpl^2\left({V_\phi''\over V_\phi}\right) 
\eea
The number of e-foldings is given by 
\beq
N_e = {1\over\mpl}\int^{\phi_*}_{\phi_c}{d\phi\over\sqrt{2\,\epsilon(\phi)}}
\label{NeDef}\eeq
Here $\phi_c = v\sqrt{\lambda/ g}$ is the critical value of $\phi$ where the Higgs at $h=0$ becomes unstable, and the transition occurs.  $\phi_*$ is the value of the field when modes leave the horizon that will later re-enter the apparent horizon on the scale of interest.  Here we will mainly be interested in scales accessed by low to moderate multipoles in the cosmic microwave background radiation. 

The predictions for the scalar amplitude $A_s^2$, scalar spectral index $n_s$, 
tensor-to-scalar ratio $r$, 
are given by
\bea
A_s^2 &=& {V_\phi\over 24\pi^2\mpl^4\epsilon(\phi_*)}\\
n_s &=& 1-6\,\epsilon(\phi_*)+2\,\eta(\phi_*)\\
r&=&16\,\epsilon(\phi_*)
\eea
A suite of recent observations have determined some of these parameters and constrained others \cite{WMAP,Planck,BICEP}. 
The scalar amplitude is measured to be $A_s^2\approx 2.45\times 10^{-9}$.  Using this value, we  parameterize the mass scale driving inflation as
\beq
V_\phi^{1/4}\approx 2\times 10^{16}\,\mbox{GeV}\left(r\over 0.15\right)^{1/4}
\eeq
Unless $r \lll 1$, this will be similar to the unification mass scale.
In most models of inflation that approximate agreement appears accidental.   Here our guiding principle is to realize it as an intrinsic feature.   

Recent CMB measurements yield estimates for the spectral index $n_s$ and tensor-to-scalar ratio $r$ \cite{WMAP,Planck}. 
The spectral index is measured to be $n_s=0.9603\pm 0.0073$. 
The value of $r$ is presently unknown, though constrained.  The Planck experiment \cite{Planck} claims a limit $r<0.11$
at 95\% confidence level.  The BICEP2 experiment \cite{BICEP} claimed detection, with
$r=0.2^{+0.07}_{-0.05}$, though it is probably best explained by dust.

\section{Simple Potentials}\label{Potential}

As the simplest starting point, we consider a monomial potential
\beq
U(\phi)={\ms^{4-p}\over p}|\phi|^p
\eeq
where $\gamma$ has the dimensions of energy.  
For reasons that will appear shortly, we will be especially interested in the range $0<p<2$.  The behavior of these potentials as $\phi\to 0$ is non-analytic.  We will regulate that behavior in Section \ref{Modified}.

The first two slow-roll parameters are easily computed to be
\bea
\epsilon&=&{\mpl^2\over2}{\ms^{8-2p}|\phi|^{2p-2}\over (V_0+\ms^{4-p}|\phi|^p/p)^2}\\
\eta&=&\mpl^2{(p-1)\ms^{4-p}|\phi|^{p-2}\over V_0+\ms^{4-p}|\phi|^p/p}
\eea
Let us provisionally assume that inflation ends for $\phi_c\ll\phi_*$ (we will return to this point in Section \ref{Modified}).
Again recalling that we are interested in the regime: $0<p<2$, we can neglect terms that depend on $\phi_c$ here, and find that the number of e-foldings is given by
\beq
N_e = {V_0\,|\phi_*|^{2-p}\over\ms^{4-p}\mpl^2}\left({1\over 2-p}+{\ms^{4-p}|\phi_*|^p\over 2 p V_0}\right)
\label{Ne}\eeq

It is useful to define three dimensionless quantities $\phi_d\equiv{|\phi_*|\over\mpl}, \ms_d\equiv {\ms^{4-p}\mpl^p\over V_0}, u_d\equiv{\ms_d\phi_d^p\over p} = {U(\phi_*)\over V_0}$.  $u_d<1$ if the GUT vacuum energy $V_0$ dominates during inflation.

$\epsilon$ and $\eta$ can be written in terms of $N_e$ and $u_d$ as
\bea
\epsilon(N_e,u_d) &=& {p \, u_d \over 2 N_e (1+u_d)^2}\left({1\over 2-p}+{u_d\over2}\right)\\
\eta(N_e,u_d) &=&  {p-1 \over N_e (1+u_d)}\left({1\over 2-p}+{u_d\over2}\right)
\eea
Using $n_s=1-6\epsilon+2\eta$ and $r=16\epsilon$, we obtain expressions for $n_s=n_s(N_e,u_d)$ and $r=r(N_e,u_d)$.
We also find negligible running of the spectral index.

\subsection{Results}

As we scan $u_d$ over the range $0<u_d<\infty$, for fixed values of $p$ and $N_e$, a curve in the $r$ vs $n_s$ plane is mapped out. This is presented in Figure \ref{fig:nsvsr}. The usual range for the number of e-foldings is $50<N_e<60$, so we have picked a representative value of $N_e=55$ in the solid curves. A much lower value of $N_e=35$ is given by the dashed curves, whose justification will be given in Section \ref{Modified} when we alter the shape of the potential near $\phi=0$.

In the figure, blue curves are for $p=1$, red curves are for $p=2/3$, and green curves are for $p=3/2$. Note that for higher values of $p$ there is a stronger tendency towards larger $n_s$ for a fixed value of $r$. Hence this heads towards a blue spectrum, when the data clearly prefers a red spectrum. This is the reason why we are not interested in $p\geq 2$ and have restricted our attention to the range $0<p<2$. (In fact for $p\geq 2$ there is a divergence in the number of e-foldings if we take $\phi_c\to 0$.)

Now lets evaluate the spectral index and tensor-to-scalar ratio in a couple of limits:
\bea
&&n_s\approx 1-{2(1-p)\over N_e(2-p)},\,\,\, r\approx{8\,p\, u_d\over N_e(2-p)};\,\,\,\, u_d\ll 1 \\
&&n_s\approx 1-{2+p\over 2N_e},\,\,\,\,\,\,\,\,\,\,\,\,\,\,\, r\approx{4 p \over N_e};\,\,\,\,\,\,\,\,\,\,\,\,\,\,\,\,\,\,\,\,\,\, u_d\gg 1 
\eea
In order to be $V_0$ dominated during inflation, we require
$u_d < 1$.
This implies that the tensor-to-scalar ratio is bounded by
\beq
r < r_{\bigcirc} =  {p(4-p)\over N_e(2-p)}
\label{bound}\eeq
This upper limit $r_{\bigcirc}$ is indicated in Figure \ref{fig:nsvsr} by the circles for each choice of $N_e$ and $p$.
For example, for $N_e=55$ and $p=1$ (solid blue), this corresponds to $r<0.06$; which is certainly consistent with WMAP \cite{WMAP} and Planck \cite{Planck} data. In Section \ref{Modified} we will see that it may be reasonable to consider lower values of $N_e$, which would raise $r$ and lead to easier detection.

\subsection{Planck-Scale Field Value}

When $\phi_*$ significantly exceeds the Planck scale, neglect of quantum gravitational corrections is hard to justify.  Models in which we avoid that issue are therefore favored, other things being equal. 
To bring this feature out it is useful to express $n_s$ and $r$ as functions of $N_e$ and $\phi_d$. To do so, we need to express $\ms_d$ in terms of these variables also.  

$\epsilon$ and $\eta$ (and through them $n_s$ and $r$) can be written in terms of $N_e$ and $\phi_d$ as
\bea
\epsilon(N_e,\phi_d) &=& {2\phi_d^2\over(2N_e(2-p)+\phi_d^2)^2}\\
\eta(N_e,\phi_d) &=& {2(p-1)\over 2N_e(2-p)+\phi_d^2}
\eea

As a concrete illustration, let us take
$\phi_* = 1\sqrt{G_N} = \sqrt{8\pi}\,\mpl$,
which is the ordinary Planck mass.  Then
\beq
r_{\square} = {64\pi\over (N_e(2-p)+4\pi)^2}
\eeq
These values $r_{\square}$ are indicated in Figure \ref{fig:nsvsr} by the squares for each choice of $N_e$ and $p$.
For $N_e=55$ and $p=1$ (solid blue) we find to $r\approx0.05$, consistent with the bound $r_{\bigcirc}\approx 0.06$ quoted in (\ref{bound}).   This value or $r$ is compatible with the Planck results \cite{Planck}.
For the scalar spectral index we have
\beq
n_{s\square} = 1-{2N_e(2-p)(1-p) + 8(4-p)\pi\over (N_e(2-p)+4\pi)^2}
\eeq
For $N_e=55$ and $p\sim 1$ this gives $n_s\approx 0.98$, which is slightly higher than the central value of $n_s\approx 0.96$ preferred by Planck data \cite{Planck}, but not excluded. 

In Section \ref{Modified} we argue that lower values of $N_e$, which lead to larger $r$ and smaller $n_s$, are not implausible.

Using the full expression for $V_\phi$ and $\epsilon$, the prediction for the amplitude of scalar modes is 
\beq
A_s^2 = {p(2N_e(2-p)+\phi_d^2)^3\over 48\pi^2(2-p)\phi_d^2(2pN_e-\phi_d^2)}{V_0\over\mpl^4}
\eeq
Using our reference values $N_e\sim 55$, $p\sim1$, and $\phi_d\sim\sqrt{8\pi}$, we find that the observed value of the scalar amplitude ($A_s^2\approx2.45\times 10^{-9}$) corresponds to $V_0\approx (1.4\times 10^{16}\,\mbox{GeV})^4$ which is close to the GUT energy density.  Running the logic backwards, we see that by putting $V_0$ at the GUT scale and $\phi_*$ at the Planck scale, we are led to predict roughly the correct scalar mode amplitude.

\section{Regulated Potentials}\label{Modified}

In the previous section we considered the potential $U(\phi)\propto|\phi|^p$. This potential is not smooth near $\phi\sim 0$, so it is unusual from the point of view of particle physics. Ordinarily we would prefer a potential that is smooth and can be expanded as a regular tower of operators. Fortunately, only the monomial behavior at large $\phi$  is important for the fluctuations that appear in the CMB.  

Small $\phi$ does, however, enter into determining  the total number of e-foldings.  
To assess the modification that may arise for a regulated potential, let us consider
\beq
U(\phi) = {\ms^{4-p}\lm^p\over p}\left(\left(1+{\phi^2\over\lm^2}\right)^{\!{p\over2}}-1\right)
\label{monodromy}\eeq
This is the sort of function that appears in axion monodromy models \cite{McAllister:2008hb,ModMcAllister}, with $p=1$ or $p=2/3$ as possibilities.  (Though axions, with an approximate shift symmetry, will not support the tree-level coupling to $h$ assumed in eq.~(\ref{FullPot}).)  $\Lambda$ is a microphysical parameter that sets the scale for the crossover from quadratic to the $p$-power behavior of the potential.
The specific details of how $U(\phi)$ interpolates between $U\propto \phi^2$ at small $\phi$ and $U\propto |\phi|^p$ at large $\phi$ are not crucial to our main consideration, which we will now explain.

We can split the integral in (\ref{NeDef}) expressing the number of e-foldings into two pieces 
$N_{tot} = N_e + \Delta N_e$ where 
\beq
N_e = {1\over\mpl}\int^{\phi_*}_{\lm}{d\phi\over\sqrt{2\,\epsilon(\phi)}},\,\,
\Delta N = {1\over\mpl}\int^{\lm}_{\phi_c}{d\phi\over\sqrt{2\,\epsilon(\phi)}}
\eeq
In the first term ($\phi>\lm$) the monomial approximation is appropriate. Assuming $\phi_*\gg\lm$, this leads back to our preceding analysis.  So we can approximate $N_e$ by the expression given in eq.~(\ref{Ne}). 

In the second term ($\phi<\lm$) the quadratic approximation is appropriate. Then
\beq
\Delta N\approx{\lm^{2-p} V_0\over \ms^{4-p}\mpl^2}\log\!\left(\lm\over\phi_c\right)
\label{dNe}\eeq
This can easily be a numerically significant contribution.  For example, 
with $\phi_d=\sqrt{8\pi}$, $p=1$, and $N_e=55$ we find $\gamma_d\approx 0.13$, so with $\Lambda\sim\mpl$ the prefactor of the logarithm is $\mathcal{O}(10)$.   

Let us consider, for example, $\Delta N\sim 20$. Then we only require $N_e\sim 35$, rather than $\sim 55$. 
The case of $N_e=35$ is given by the dashed curves in Figure \ref{fig:nsvsr}. 
Note that on these dashed curves ($N_e=35$) the squares sit slightly higher than the circles. This means that when $\phi_*=1/\sqrt{G_N}$ the ratio of $U(\phi_*)$ to $V_0$ is slightly larger than 1.  In for $V_0$ to dominate, we should consider $\phi_*$ slightly smaller than $1/\sqrt{G_N}$.

For the consistency check of this analysis, we need $\phi_c<\lm$.  If we take $\lm\sim\mpl$, $\lambda\sim 1$, $v\sim 10^{16}\,$GeV, then from $\phi_c = v\sqrt{\lambda/ g}$ this condition is satisfied for $g\gtrsim 10^{-4}$.  A coupling of this order generates radiative corrections to the quartic potential that are roughly comparable to the quartic coupling that arises from expanding eq.~(\ref{monodromy}).  

\bigskip

\section{Discussion}\label{Conclusions}

We have demonstrated that two attractive ideas, {\it viz}. (i) vacuum energy associated with unification drives inflation, and (ii) the inflaton field displacement is of order the Planck scale, can be incorporated into a class of phenomenologically viable inflation models.  

Because the field displacement of $\phi$ generally approaches the Planck scale, these models are sensitive to quantum gravitational effects and other corrections.  For discussion of these issues, see, e.g., \cite{Baumann:2009ds,Hertzberg:2011rc,Hertzberg:2014aha,Baumann:2014nda}.  

Our treatment of the unified symmetry and its breaking has been extremely schematic.   This seems justified strategically, since possible cosmological implications of the overall unification scale raise different issues from, and can be discussed independently of, such issues as dark matter, baryogenesis, and topological defects, which depend on many underdetermined details.  For some relevant ideas, see \cite{Hertzberg:2013jba,Hertzberg:2013mba}.

One comment however, does seem in order, since it touches on a major historical motivation for inflation, that is the removal of magnetic monopoles (which accompany virtually all modern attempts at unification) as a cosmological relic.   If our $h$ field simply encoded the magnitude of an adjoint of $SU(5)$, involved in breaking the unified symmetry $SU(5) \rightarrow  SU(3) \times SU(2) \times U(1)$, for example, then its condensation {\it after\/} inflation would render inflation irrelevant as a dilution mechanism, leaving an unacceptable relic monopole density.   An attractive alternative may be to start with $SO(10)$, and have $SO(10) \rightarrow SU(5) \times U(1) $ in the pre-inflationary phase.   Further breaking of this symmetry need not lead to monopole production, any more than does the much later $SU(2) \times U(1) \rightarrow U(1)$ electroweak symmetry breaking.

\begin{center}
{\bf Acknowledgments}
\end{center}

This work has been supported by the Center for Theoretical Physics at MIT and by 
the U.S. Department of Energy under cooperative research agreement Contract Number DE-FG02-05ER41360.


\end{document}